\documentclass{article}

\usepackage{arxiv}

\usepackage[utf8]{inputenc} 
\usepackage[T1]{fontenc}    
\usepackage{hyperref}       
\usepackage{url}            
\usepackage{booktabs}       
\usepackage{amsfonts}       
\usepackage{nicefrac}       
\usepackage{microtype}      
\usepackage{lipsum}
\usepackage{graphicx}
\usepackage{natbib}
\title{The Thousand Brains Theory 2.0: An Extension for the Long-Range Connections of the Neocortical Heterarchy}

\author{
 Jeff Hawkins \\
   Thousand Brains Project\\
  Santa Rosa, CA, USA\\
  \texttt{jhawkins@thousandbrains.org} \\
   \And
 Niels Leadholm \\
   Thousand Brains Project\\
  Santa Rosa, CA, USA\\
  \texttt{nleadholm@thousandbrains.org} \\
  \And
 Viviane Clay \\
  Thousand Brains Project\\
  Santa Rosa, CA, USA\\
  \texttt{vclay@thousandbrains.org} \\
}

\author{
      \href{}{\hspace{1mm}Jeff ~Hawkins},
      \href{}{\hspace{1mm}Niels ~Leadholm},
      \href{}{\hspace{1mm}Viviane ~Clay}\\
      Thousand Brains Project, Santa Rosa, CA, United States \\
      \texttt{\{jhawkins, nleadholm, vclay\}@thousandbrains.org}
    }
    
\begin{document}
\begingroup
\renewcommand{\today}{%
  \shortstack{
    First Submitted July 9, 2025\\
    Revised August 20, 2026
  }%
}
\maketitle
\endgroup
\begin{abstract}
Vernon Mountcastle hypothesized that the basis for intelligence in mammals is the replication of a general computational unit, the cortical column. The Thousand Brains Theory proposed that each column is a sensorimotor system, capable of learning structured models of objects by integrating sensory input over multiple movements. Previous papers on the Thousand Brains Theory focused on the computations that occur within individual cortical columns, and how columns can use long-range connections to rapidly reach a consensus. However, several prominent long-range connection types in the neocortex were not addressed by the theory. These include hierarchical feedforward and feedback connections, as well as those that go through the thalamus. In addition, several theoretical requirements were not addressed. These include how the cortex learns compositional objects, and how information is converted from the egocentric perspective of sensors to the allocentric perspective of models in the cortex. In this paper, we extend the Thousand Brains Theory to address these issues.

We begin by reviewing the anatomy of long-range neocortical connections, arguing that they form a \textit{heterarchy}, rather than hierarchy, which has made their functions challenging to understand through existing theoretical models. We then propose specific roles for each of these connections. First, the thalamus converts the orientation of features and movement information from an egocentric perspective to the allocentric perspective of learned models. Second, hierarchical feedforward, feedback, and cortico-thalamo-cortical projections enable columns to learn compositional models. We discuss the relationship of our proposals to experimental findings at the levels of anatomy, neurophysiology, and behavior, along with testable predictions for future experimental work. We conclude by discussing the implications of the extended Thousand Brains Theory for the design of artificial intelligence.
\end{abstract}


Among the billions of neurons that form the neocortex, along with their complex connectivity, one can observe several structural principles. When looking at a cross-section anywhere in the neocortex, a layered structure is observed, with different layers having different compositions of neuron types, densities, and response properties \citep{cajal1899comparative, Gilbert1977LaminarCortex}. Orthogonal to these layers, neurons can be grouped into functional units, called cortical columns, with a diameter of around 300-600 $\mu$m. The concept of columns came from the observation that all the neurons within such an area respond to input from the same sensory patch. Neurons in an adjacent column respond to a different or partially overlapping sensory patch \citep{Mountcastle1997TheNeocortex}. Columns have been proposed to implement a canonical circuit that is repeated across the neocortex \citep{mountcastle1957, douglas_neuronal_2004, markram_interneurons_2004, da_costa_2010}. 

All neurons within a cortical column connect to other neurons in the same column. Some neurons also send a branch of their axon outside of the column. These are the long-range connections that this paper focuses on. There are two types of long-range connections that we consider. Cortico-cortical (CC) connections originate in one part of the neocortex and terminate in another part of the neocortex. Cortico-Thalamo-Cortical (CTC) connections originate in the neocortex and project to the thalamus, which then sends axons back to the neocortex. Both CC and CTC connections remain to be fully understood.

Theoretical models of the neocortex typically emphasize its hierarchical nature \citep{Felleman1991DistributedCortex, DiCarlo2012HowRecognition, markov2014, rolls2016, Grossberg2021, Rolls2021, rao2024sensory}, focusing on the role of feedforward \citep{Hubel1959ReceptiveCortex, Fukushima1980Neocognitron:Position, Riesenhuber1999HierarchicalCortex} and feedback connections \citep{Rao1999, schwabe_role_2006, lillicrap_backpropagation_2020} in enabling object-like representations given a sufficiently deep hierarchy. In this view, sensory information flows up a hierarchy of regions, with each level processing increasingly complex features. Complete objects are only recognized after several levels of processing. Models that consider the role of non-hierarchical, lateral connections \citep{iyer_contextual_2020, Pogodin2021TowardsNetworks} still emphasize a strict hierarchical form of representations, where early layers encode only simple stimuli. Despite this emphasis in the theoretical literature, ample experimental evidence supports a more complex view. Columns in both "low-level" and "high-level" cortex communicate directly with sensory and motor regions, are able to respond in parallel rather than just sequentially, and communicate with other columns even when they share no clear hierarchical relation \citep{HedgeFelleman2007, voges_2010, henschke_2015, usrey2019, suzuki_2023}. For these reasons, we believe that neocortical organization is best described as a \textit{heterarchy} rather than a hierarchy. While this discrepancy between theory and experimental findings is increasingly recognized \citep{suzuki_2023}, a unified model to understand the nature and benefits of heterarchical processing remains outstanding.

The Thousand Brains Theory (TBT) proposed that all cortical columns implement a sensorimotor computational unit that learns structured representations of objects by integrating movement in reference frames. The original theory provided detailed computational models for the role of individual columns, as well as a role for some non-hierarchical connections in enabling columns to rapidly reach consensus during inference \citep{Hawkins2017, Hawkins2019ANeocortex, hawkins2021thousand}. However, the original TBT did not provide detailed proposals for the purpose of hierarchical feedforward and feedback connections, nor did it explain the purpose of the thalamus in cortical computations.

To address these limitations in prior work, we first review the anatomy of the key long-range connections in the neocortex. We then summarize the Thousand Brains Theory to date, emphasizing the importance of reference frames and sensorimotor learning in all columns, as these are fundamental to understanding the purpose of long-range connections. We then extend the Thousand Brains Theory, making the following contributions:
\begin{itemize}
    \item We propose that thalamic relay cells transform the orientation of sensory and movement data to compensate for changes in the egocentric orientation of moving sensors and the allocentric orientation of models in the cortex. They perform a similar orientation transform for CTC connections. We provide a circuit-level proposal for how such computations could be carried out in the thalamus.
    \item We propose that the purpose of hierarchical feedforward and feedback Cortico-Cortical (CC) connections is to mediate the learning of compositional objects, that is, they define parent-child relations between objects. Importantly, both parent and child representations are whole objects, rather than simple feature primitives. Hierarchical feedback connections support learning specific spatial relations between child and parent objects.
\end{itemize}

These proposals represent a fundamentally new basis for understanding the purpose of long-range connections in the neocortex, as well as a novel proposal for the role of the thalamus. At a computational level, these proposals can explain the robust sensorimotor capabilities seen in mammals, including the ability to recognize objects despite changes in their orientation or the orientation of sensors, as well as the ability to rapidly learn and infer compositional objects. We conclude by discussing the implications of the extended Thousand Brains Theory for the design of artificial intelligence.

\section{Long-Range Connections in the Neocortex}

We define “long-range” connections as axonal projections that travel outside of the cortical column where they originate. This includes axons that enter the white matter and reenter the cortex elsewhere, axons that project to the thalamus, and axons that travel horizontally within the cortex for distances of 1mm or more. These connections all play a role in how columns communicate with one another, either directly, or indirectly via the thalamus.

We do not include all known projections from the cortex. For example, we do not discuss projections to the striatum or claustrum. These are undoubtedly important, but they fall outside the scope of this paper. Intracolumnar connections, that is axons that originate and terminate within the bounds of a single cortical column, are also not the subject of this paper, although we will discuss some intracolumnar connections as necessary.

There are many empirical studies about long-range connections. Reported findings differ by species, by sensory modality, and by experimental methods. This makes it difficult to provide an accurate summary that applies to all mammals in all situations. In the following sections, we attempt to summarize broad classes of connections that exist in most mammals. Some details will vary by species and sensory modality. However, this does not prevent us from proposing functional roles for the different connection types. For simplicity, and due to the extent to which they have been studied, we will also focus on primary and secondary sensory regions such as V1 and V2 as examples. However, we believe that the same general principles apply across the entire neocortex.

We divide the long-range cortical pathways into four groups:
\begin{enumerate}
    \item Hierarchical Cortico-cortical Connections
    \item Non-hierarchical Cortico-cortical Connections
    \item Cortico-thalamic Connections
    \item Layer 5 Motor Connections
\end{enumerate}
The latter two types of connections have both hierarchical and non-hierarchical properties. This complex mixture of hierarchical and non-hierarchical connectivity can be described as a heterarchy, and we argue that this has made understanding the purpose of these long-range connections challenging. In this paper, we will demonstrate how the Thousand Brains Theory represents a unique framework for addressing this gap in understanding.

\subsection*{Hierarchical Cortico-Cortical Connections}
\label{sec:hierarchicalCC}

\begin{figure}[!htbp]
    \centering
    \includegraphics[width=1.0\textwidth]{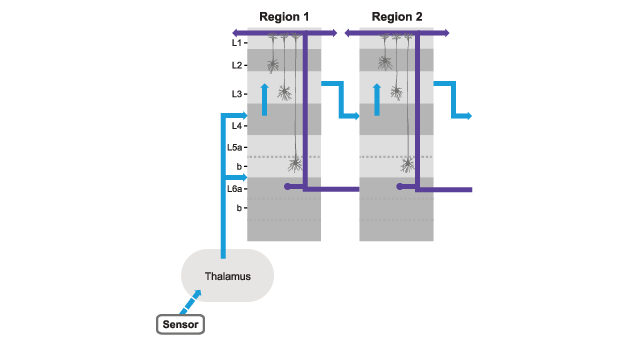}
    \caption{Classic feedforward (blue) and feedback (purple) pathways illustrated on two cortical columns. In this and subsequent figures, the rectangles represent individual cortical columns, not complete regions. The two columns are located in different regions in a hierarchical sensory-processing stream (for example, V1 and V2). Connections between regions typically exit a column via layer 6 (L6), travel some distance through the white matter, and reenter the cortex via L6, where they then ascend to their destination layer in the receiving column. In this and the following figures we do not show this detail.}
    \label{fig:hierarchical_c}
\end{figure}

The classical cortical hierarchy is defined by an asymmetry in laminar connectivity between regions. More precisely, projections from a lower region to a higher region connect to different cellular layers than the reverse projections from the higher region to the lower region \citep{Felleman1991DistributedCortex, rockland1979, burkhalter_1989}. These classical feedforward and feedback pathways are illustrated in figure \ref{fig:hierarchical_c}.

In the feedforward direction, neurons carrying information from a sensor project to relay cells in first-order (FO) nuclei of the thalamus. The relay cells then project to layer 4 (L4) in the first region in a hierarchy of regions \citep{usrey2019}. The L4 cells project to cells in L3 immediately above L4 in the same column. L3 cells are considered a primary output of a column, as they project to L4 in the next region of the hierarchy \citep{hirsch_2006, harris_2013}. This pattern is repeated in a chain of regions. This pathway is considered hierarchical because the connections are asymmetric; we do not see L3 cells in region 2 projecting back to L4 cells in region 1 \citep{Felleman1991DistributedCortex}.

Notice that relay cells in the thalamus do not only project to L4 as is sometimes reported. There are also distinct projections from the thalamus to the border between L5 and L6 \citep{oberlaender_2012, constantinople_2013, harris_2013}.

The primary feedback pathway is shown in purple. This pathway originates in L6a. An L6a cell axon enters the white matter and then reenters the neocortex in the next lower region of the hierarchy. As the axon rises through the cortical layers, it may have short collaterals and exhibit clustered boutons in deeper layers, before reaching L1 \citep{rockland_1989, rockland1996collateralized, markov2014,shipp2007,rockland_what_2019}. In L1, the axon spreads horizontally over multiple cortical columns \citep{rockland_1989, stettler_2002}. Neurons with cell bodies in L2, L3, and L5b (shown in light gray) send apical dendrites toward L1 where they may form synapses with the feedback axon \citep{harris_2013, schuman_2021}. The broad horizontal spread in L1 contrasts with the short collaterals in the deeper layers, which are narrowly focused.

\subsection{Non-Hierarchical Connections}
\begin{figure}[!htbp]
    \centering
    \includegraphics[width=\textwidth]{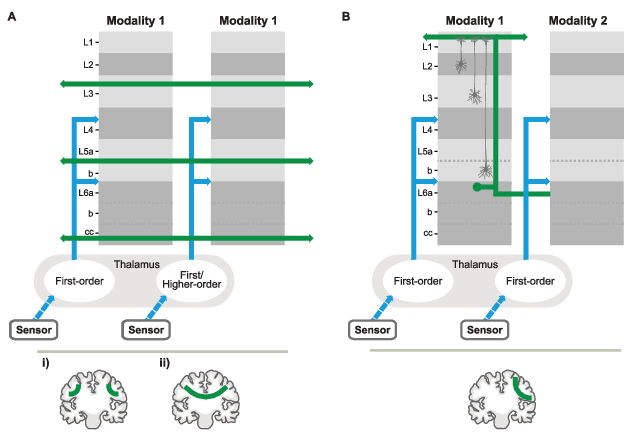}
    \caption{There are numerous long-range connections that do not make sense in a classical hierarchy. A) Cells in layers 3, 5, and 6 connect to the same layers in other columns (green). These connect columns within the same region (i), and between contralateral sides of the brain (ii). B) Columns in different modalities, such as V1, S1, and A1, also share connections. These connections have a similar anatomical structure to the hierarchical feedback connections in figure \ref{fig:hierarchical_c}, yet they are reciprocal and don't make sense in any classical view of hierarchy. In addition, all columns (A and B) receive driving input from thalamic relay cells (shown in blue). For example, columns in both V1 and V2 receive input from the retina, suggesting that V1 and V2 can act in parallel, not just hierarchically.}
    \label{fig:nonH_c}
\end{figure}

Figure \ref{fig:nonH_c} illustrates long-range cortico-cortical connections that are not hierarchical. On the left side of the figure, neurons in layers 3, 5, and a subset of layer 6 send their axons long distances where they form synapses with columns in the same region \citep{gilbert_columnar_1989, angelucci_circuits_2002, voges_2010} and with columns on the opposite side of the brain (e.g., somatic regions representing the left and right hands) \citep{Thomson2010NeocorticalReview, payne_1991, fame_2011}. Typically, these cells will form connections with cells in the same layer as the originating cell.

Direct connections also exist between regions in different sensory modalities, including primary regions such as V1, A1, and S1 \citep{henschke_2015}. These connections can resemble the feedback connections mentioned earlier in that they originate in L6 of one region and terminate in L1 and L6 of the recipient region \citep{rockland_2003, falchier_2010, smith_2010}. However, these connections exist between primary sensory regions, and are known to be reciprocal between regions \citep{henschke_2015}. As such, they cannot be interpreted as performing hierarchical computations.

Although sensory input enters the neocortex through primary thalamic nuclei, such as LGN, and then goes from region to region via hierarchical connections as shown in figure \ref{fig:hierarchical_c}, high-order regions also receive direct sensory or other sub-cortical input via high-order thalamic nuclei  \citep{bullier_1983, glickstein_1969, warner_2010, lopez_2011, suzuki_2023} as shown in blue in figure \ref{fig:nonH_c}A. The connections shown in this figure are consistent with regions being able to operate in parallel, not only in a sequential hierarchy.

\subsection{Cortico-Thalamic Connections}
\begin{figure}[!htbp]
    \centering
    \includegraphics[width=1.0\textwidth]{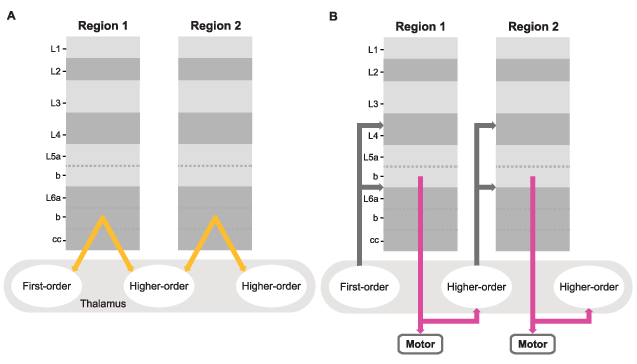}
    \caption{Cortico-thalamic and motor connections appear to be both hierarchical and non-hierarchical. A) Layers 6a and 6b project back to the thalamic nucleus that provides input to the column. Layer 6b also projects forward to the next higher-order thalamic nucleus. Evidence indicates that layer 6 cortico-thalamic connections do not directly cause thalamic relay cells to fire. Instead, they modify how relay cells respond to subcortical input. B) Every column has a motor output, originating in L5b, which exits the neocortex. A copy of this output is routed through the thalamus to the next cortical column.}
    \label{fig:cortico_thl_c}
\end{figure}

The neocortex and thalamus are intimately connected. All sub-cortical driving input to the neocortex passes through thalamic relay cells as illustrated in figure \ref{fig:hierarchical_c} and \ref{fig:nonH_c}. However, there are many more projections from the neocortex back to the thalamus than there are driving inputs to thalamic relay cells \citep{sherman_chapter4}. Most of these projections originate in layer 6 as illustrated in figure \ref{fig:cortico_thl_c}A. Despite the large number of these connections, it is believed that they play a modulatory role and do not directly cause relay cells to fire \citep{sherman_2005}.

The primary set of cortico-thalamic projections we will discuss, shown in yellow, originates in L6b. L6b cells project back to relay cells that send input to the column. In addition, L6b cells project forward to the next higher-order thalamic nucleus \citep{hoerder_suabedissen_subset_2018,Thomson2010NeocorticalReview, Sherman_guillery_book_2013}. Thus, most thalamic nuclei will receive converging L6b input from two hierarchically arranged columns.

Whether L6 cortico-thalamic projections are hierarchical is debatable. On the one hand, they project forward and backward to thalamic nuclei. On the other hand, they only seem capable of modulating the firing of relay cells, and they appear to send the same information to both lower and higher levels of the hierarchy.

\subsection{L5 Motor Connections}
The final long-range connections we discuss are the motor outputs of the neocortex, shown in pink in figure \ref{fig:cortico_thl_c}B. The motor output originates in the large intrinsically-bursting cells in L5b \citep{usrey2019}. These cells exit the neocortex and project to motor-related subcortical areas of the brain. For example, L5b cells in visual regions of the neocortex project to the superior colliculus, which integrates inputs from multiple modalities and generates eye movements \citep{liu_2022}. L5b cells in somatic/motor cortex project to the spinal cord to control limbs \citep{frezel_2020}. As far as is known, every column in the neocortex has a motor-related output originating in L5b \citep{usrey2019}.

As shown in figure \ref{fig:cortico_thl_c}B, L5b axons bifurcate and send an axon branch to the next hierarchically higher region via the thalamus. It is believed that these connections are strong enough to activate relay cells and thus drive input to the next region via high-order thalamic nuclei \citep{Sherman_guillery_book_2013, usrey2019}. Since L5b cells represent a motor output, the projection to the next region is a feedforward efference copy of a motor command.

These connections suggest that motor processing and motor output are an intrinsic part of every cortical column. They also suggest that the motor output of the neocortex operates in both hierarchical and non-hierarchical ways. When columns in multiple regions, such as V1 and V2, directly send motor signals to subcortical areas, they are acting in parallel, non-hierarchically. When V1 sends a copy of its motor signal to V2, via the thalamus, information flows in a serial, hierarchical fashion.

\section{The Thousand Brains Theory 1.0 Reviewed}

Despite a wealth of empirical data, there are no comprehensive theories that explain the function of the various long-range connections described in the previous section. We propose that these connections can all be understood if we conceptualize every cortical column as a sensorimotor system, which is the core idea of the Thousand Brains Theory (TBT) \citep{Hawkins2019ANeocortex, hawkins2021thousand}. We now summarize the existing TBT before extending it to include the connections described earlier.

\subsection{A Canonical, Sensorimotor Microcircuit}\label{sec:microcircuit}

The brain must keep track of how its sensors are moving to understand its changing sensory input. For example, the somatosensory cortex needs to be informed of how the individual parts of the hands and body are moving to understand and predict the next tactile sensations. Similarly, visual regions cannot understand or predict the next retinal input without knowing how the eyes are moving. The need to integrate movement data with sensation has been known for over one hundred years, but it is unclear where and how motion data enters the neocortex, and how it is integrated into cortical processing. Most theories of the neocortex omit this critical role of movement, or limit motor processing to circumscribed regions of the neocortex.

The TBT proposed that sensorimotor integration occurs in every cortical column. In particular, each column receives sensory data and movement data. The movement data reflects how the sensor is moving. By integrating sensory and movement information over time, each cortical column is able to learn structured models of the world. For example, a column receiving input from a fingertip can learn a 3D model of an object by keeping track of where the finger is as it moves over the edges and surfaces of the object. Models of any particular object exist in many columns, including columns in different sensory modalities.

\begin{figure}[!htbp]
    \centering
    \includegraphics[width=1.0\textwidth]{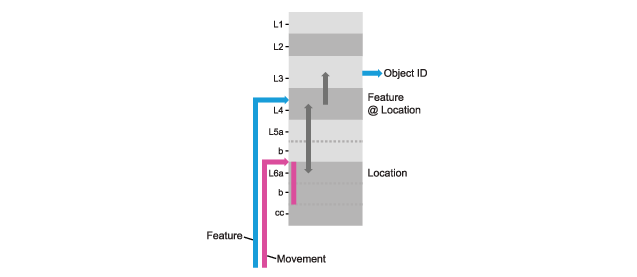}
    \caption{The core sensorimotor circuit of a cortical column per the Thousand Brains Theory. Cortical columns integrate sensory features and movement information to learn and recognize models. Sensory features in L4 are bound to locations in L6a via bidirectional associative connections. As a sensor moves, new features are represented in L4, while the location represented in L6a is updated using the incoming movement information. L3 pools over multiple activations in L4 and represents the object being sensed.}
    \label{fig:column}
\end{figure}

Figure \ref{fig:column} shows how a single cortical column implements sensorimotor processing per the original TBT. By integrating feature and movement data, the column learns structured models of the world. This requires the column to know the location of its associated sensor patch relative to the object being sensed. According to the TBT, a column represents the location of its sensor patch using cells similar to grid cells, which were first found in the entorhinal cortex \citep{Hafting2005MicrostructureCortex}. Grid cells represent an animal's location relative to an environment's reference frame, whereas the TBT proposes that analogous cells in cortical columns represent a sensor patch's location relative to an object's reference frame. The incoming movement information, indicated by the pink line in figure \ref{fig:column}, updates the representations of location via path integration \footnote{The prediction that cortical columns use reference frames to model objects remains a cornerstone of the original TBT, as well as the TBT 2.0. As such, we discuss biological evidence around this proposal, including the specific prediction that grid cells might exist in L6, in the Appendix (\ref{sec:appendix_grid_cells}).}.

The TBT proposed that cells in L6a represent the sensor's location on the object. When a feature is sensed, it arrives in L4, where it is paired with the feature's location. This is achieved by the bidirectional associative connections between L4 and L6a (gray arrow) \citep{Thomson2010NeocorticalReview}. L4 forms a neural representation that is unique to both the feature and its location on the object. L3 performs a pooling operation over multiple activations in L4. Thus, L3 is a stable representation of the object being observed, while L4 and L6a change with each movement of the sensor patch. In this way, a column will learn models of objects that are larger than what the column’s sensor patch can observe at once.

\subsection{Non-Hierarchical Connections: Voting and Multisensory Binding}

In addition to the above microcircuit, the TBT proposed that the long-range, non-hierarchical connections (green lines in figure \ref{fig:nonH_c}) enable columns to rapidly reach a consensus during inference, and unite inputs from multiple modalities into a singular percept. Typically, many columns will be observing the same object at the same time. For example, multiple columns in V1 may detect visual features on an object, albeit at different locations. At the same time, multiple patches of skin may touch the object, again, at different locations. All these columns will learn models of the same underlying object. However, each column can observe only one part of the object at any point in time. On its own, a column would typically have to make a series of movements and observations to infer the object. 

The lateral connections between columns allow them to reach a consensus more quickly, often in a single visual fixation or a single grasp of the hand. L3, which represents the current object being inferred, is stable over movements. During learning, the active cells in L3 in different columns can be associatively linked via a simple Hebbian learning rule. After this associative pairing, if one or more columns know what object they are observing, they will inform other columns of what object they should be observing. If multiple columns are uncertain, then the associative pairing across columns will narrow down the possible objects to only those that are consistent with the evidence of all the columns \citep{Hawkins2017}. This process, termed `voting', enables cortical columns to recognize objects more rapidly and in the presence of noise. Importantly, voting works within and across sensory modalities.

In the classical, hierarchical view of the neocortex, primary sensory regions are only capable of representing simple sensory features. However, it is unlikely that primitive features would be sufficiently correlated in the real world that the presence of one would predict the cross-modality presence of the other. The TBT proposes that all columns represent whole objects, even in primary sensory regions, and therefore voting on objects can occur between modalities. This explains otherwise puzzling observations such as the long-range connections between S1, A1, and V1 \citep{henschke_2015}. Similarly, with objects recognized at multiple levels in the cortical hierarchy, it is no longer surprising that non-hierarchical connections are also observed connecting regions such as S1 and S2 \citep{denardo_2015}.

Non-hierarchical connections for voting represent one powerful way for columns to communicate. In the following sections, we extend the TBT to explain the role of the other long-range connections of the neocortex.

\section{The Thousand Brains Theory 2.0}

\subsection{Thalamic Transformations of Sensory and Motor Input Data}\label{sec:first_thalamus_motor}
We begin our theoretical discussion by focusing on cortico-thalamic and motor connections. We need to understand how information is transformed as it enters and exits the neocortex before we can address hierarchical connections. There are three required transformations.

The first required transformation is to change the orientation of the sensed features. Consider reading a line of text (such as this one) after rotating your head 20 degrees. While the eye performs some minor accommodation for head rotations, the vast majority of this rotation will be unaccounted for at the level of the retina \citep{Schworm2002}. Without further processing, visual features would therefore enter the cortex at atypical orientations from those experienced when learning to read. We do not perceive that the world has rotated, nor do we have difficulty recognizing the rotated text. This tells us that the orientation of sensory input is compensated for somewhere between the retina and the models in the cortex. A similar problem exists for other sensory modalities. For example, if we touch the edge of an object and then rotate our fingertip at the point of contact, the sensed pattern from the skin will change, yet our perception of the object and the object's feature remains stable.

The second required transformation is changing the orientation of incoming movement information. Sub-cortically, movement information starts in an egocentric form, such as eye movements relative to the head. However, to be useful to a cortical column, the communicated direction of the movement must be rotated to match the column's allocentric model of the object. For example, fixating at the beginning of a sentence and moving your eyes will result in different trajectories, depending on whether or not your head is tilted. Without any correction, the movement will result in an incorrect estimate of the location in the sentence. Thus, the orientation of the movement data must be rotated in the same way as the sensory data.

The third required transformation is similar but applies to the motor output of a column. Moving your eyes or fingers relative to an object requires different egocentric physical movements depending on the current orientation of the eyes or fingers relative to the object. For example, to read a line of text while the head is tilted requires moving the eyes diagonally relative to the head, whereas if the head is not tilted, the eyes need to move horizontally. A cortical column generates behaviors relative to its model of an object, that is, in an allocentric form. This must be converted to an egocentric form before action is taken.

These three transformations must be performed rapidly to minimize delays in recognizing objects and generating behaviors. This constrains where and how the transformations might occur. Another constraint is granularity. With vision, it is conceivable that a single transformation could be applied to the entire retina. However, with touch, different parts of the skin can move independently and therefore have different orientations relative to the sensed object. Thus, the transformations must be performed at a granular level. We propose that the transformations occur on a column-by-column basis.

\begin{figure}[!htbp]
    \centering
    \includegraphics[width=1.0\textwidth]{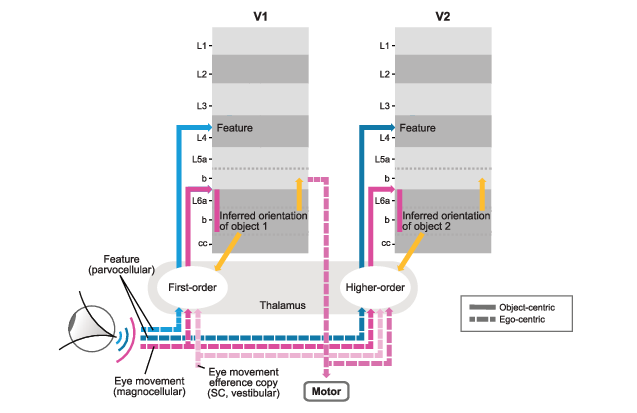}
    \caption{Feature input (blue) and movement input (pink) must be converted from a sensor's orientation (dashed lines) to an object's orientation (solid lines). We propose that this conversion occurs in the thalamus. The rotation required to transform from egocentric to object-centric space is inferred in L6b and projections from there to thalamic relay cells (yellow) modulate the relay cell activity (for more details see section \ref{sec:thalamus_mechanism}). Similarly, the motor output of a column must be converted from an object-centric orientation to an egocentric orientation. We propose that this transform occurs in L5, which also receives input from L6b. There are multiple sources of motor input to the neocortex. Input from the retina (magnocellular pathway) and the vestibular system represents observed movements. Inputs from the superior colliculus (SC) and lower-level cortical columns are efference motor copies, or movements that are about to happen. See text for full description.}
    \label{fig:motorTheory}
\end{figure}

Figure \ref{fig:motorTheory} shows our proposal for how the orientation of sensor and movement data is transformed as it enters and exits the neocortex. The key element of this proposal is that thalamic relay cells perform an orientation transform. The figure illustrates the process for vision, but the ideas apply equally to other sensory modalities. In the figure, dashed blue lines represent observed features relative to the sensor; solid blue lines represent the same features, but now rotated to be relative to the observed object. Similarly, dashed pink lines represent movements relative to the body, while solid pink lines represent movements, but now rotated to be relative to the model of the viewed object.

We propose that the egocentric to object-centric transformation of incoming sensory and motor data occurs in the thalamus, and is a primary function of thalamic relay cells. Since L6b represents the inferred orientation of egocentric space to the object's model, projections from L6b to the thalamus (shown in yellow) tell the thalamic relay cells which orientation transform is needed to make sure that sensed movement and features map correctly from egocentric into object-centric space. 

The orientation transform applied in the thalamus can be informed by both cortical and subcortical signals. Cortical signals from L6b can inform about changes in the orientation of the object in the world, such as a mug rotating in front of you. Subcortical signals can inform the thalamus of changes in the sensor's orientation, such as your head tilting relative to the mug. Both rotations can be applied using the same thalamic mechanism.

\subsubsection{Origins of Movement Data}\label{sec:origins_of_movement}
Where does the movement data come from? The source of movement data varies by modality. Figure \ref{fig:motorTheory} shows multiple sources of movement input related to vision. We propose that the magnocellular pathway is one source. Magnocellular cells in the LGN quickly respond to changing patterns on the retina and are sensitive to depth and movement-related changes \citep{livingstone_segregation_1988}. Cortical neurons in L6 that respond to this input have very broad receptive fields \citep{Gilbert1977LaminarCortex}. These attributes suggest that the magnocellular pathway allows the cortex to detect movement of the retina relative to the world. Consider that when watching someone play a first-person video game, you know if the virtual player is moving forward, backward, turning, etc. You understand how the character is moving through the virtual world as if you were controlling the character yourself, but you are not. This demonstrates that information from the retina is sufficient for determining the movement of the retina relative to the world. The magnocellular pathway is the logical source of this movement information.

Other sources of movement input for vision include the superior colliculus, which provides an efference motor copy of eye movement, and the vestibular system, which detects movements of the head and, therefore, the eyes. The spinning sensation we experience during vertigo is believed to be caused by incorrect output of the vestibular system \citep{baloh1998vertigo}. Our hypothesis that thalamic relay cells change the orientation of inputs to the neocortex explains why we see the world spinning and why our motor actions are misdirected during vertigo. In summary, the multiple sources of movement data, magnocellular, vestibular, and superior colliculus, project to the relay cells in visual areas of the thalamus.

\subsubsection{Transformations of Motor Output}\label{sec:second_thalamus_motor}

Now consider the motor output of the neocortex, which originates in L5b and projects to subcortical motor areas (figure \ref{fig:motorTheory}, pink line projecting to the "Motor" box). When a column generates movement, it is to achieve a goal relative to its models. Therefore, a column's motor output is derived in an object-centric form, yet it must be converted to an egocentric form before it can be sent to subcortical motor areas. Where does this object-centric-to-egocentric conversion occur? Unlike the inputs to the neocortex, the motor output does not pass through the thalamus on its way to subcortical motor areas. The two possible locations for changing the orientation of the motor output are in the cortical column or in the subcortical destination.

The information on what transform is needed is in L6b. L6b projects to L5 \citep{kim_2014, Thomson2010NeocorticalReview}, and we therefore propose that L5b cells perform the object-centric to egocentric orientation transform. Thus, the L5b output is in an egocentric form when it leaves the neocortex.

L5b cells also send an axon branch to the next higher cortical region. This axon branch does not project directly to cortical neurons, but instead projects to thalamic relay cells, which then project to the cortex. This is further evidence that the motor output coming from L5b is already in an egocentric form. It enters the thalamus in an egocentric form, where it is converted to the object-centric form needed to match the object being inferred in the higher region.

We highlight the heterarchical nature of the motor projection. As a direct projection from a column to motor regions, the projection is non-hierarchical, as any column can produce a motor output, not just those at the top of a hierarchy. On the other hand, the efferent copy carries information about the low-level column to the higher-level column. To enable this flexible use of the motor information, the transformations of outgoing and incoming signals that we have described are essential.

\subsubsection{A Potential Mechanism for Transforming Orientation in the Thalamus}
\label{sec:thalamus_mechanism}

We propose that the thalamus doesn’t just relay information to the cortex, but changes information from an egocentric orientation to an object-centric orientation. In this section, we propose how thalamic relay cells might perform this transformation. We use vision to describe the mechanism, but the same principles would apply to other sensory modalities.

\begin{figure}[!htbp]
    \centering
    \includegraphics[width=0.9\textwidth]{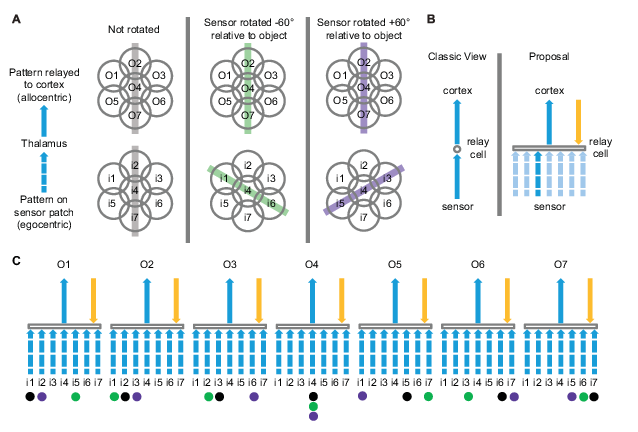}
    \caption{A thalamic circuit for transforming sensory and movement data. (A) (bottom) shows a pattern of innervation on a sensor patch at three different orientations of the sensor to the object. (A) (top) shows the pattern of innervation passed to the cortex after conversion in the thalamus. (B) and (C) show how this conversion could be achieved by relay cells in the thalamus. (B) (left) shows the classical view of a thalamic relay cell. Spikes from a single sensor cell are relayed to the cortex. (B) (right) shows our proposal that relay cells act like multiplexers, relaying one (dark blue) of several (light blue) inputs. Which input is relayed is determined by cortical or subcortical feedback (yellow). (C) shows how the conversions illustrated in (A) could be achieved using multiplexer-like relay cells. The colored dots indicate which inputs should be relayed, corresponding to the three orientation transforms shown in (A). See text for full description.}
    \label{fig:thalamusCircuit}
\end{figure}

Figure \ref{fig:thalamusCircuit} illustrates one proposal for how the thalamus could perform an orientation transform. Bear in mind that this transformation does not apply to an entire sensor array, such as the retina, or an entire thalamic region, such as LGN. The transform applies to sensor patches that project to single cortical columns. In figure \ref{fig:thalamusCircuit}A (bottom), a retinal patch is represented by seven circles labeled i1 to i7. Each circle represents a retinal ganglion cell with a center-surround receptive field. The left-most image shows a vertical line on the retina activating three ganglion cells i2, i4, and i7. Going to the right, we show how the pattern on the retina changes as the retina rotates 60 degrees to the right (green) and sixty degrees to the left (purple). The top of figure \ref{fig:thalamusCircuit}A shows seven relay cells o1 to o7. The activity of the relay cells should not change as the retina rotates relative to an observed object. 

Figure \ref{fig:thalamusCircuit}B shows two interpretations of a thalamic relay cell. The left image is the classical view. A relay cell receives input from a single retinal ganglion cell. Often, a single spike on this input axon is sufficient to cause the relay cell to spike. This is why they are called relay cells. However, relay cells are complex. They have thousands of synapses, receiving feedback from the cortical area to which they project and input from multiple subcortical areas. In addition, relay cells often form synapses with multiple retinal ganglion cells, although they appear to relay activity from only one \citep{cleland1971simultaneous, sherman_2005, hammer2015multiple}. These observations suggest that relay cells are performing a more complex function than assumed. The right image in figure \ref{fig:thalamusCircuit}B shows a proposal that relay cells act like multiplexers. The image shows a relay cell receiving seven inputs, representing seven retinal ganglion cells. The cell relays activity from one of these inputs at a time. Which input drives the relay cell depends on cortical and subcortical context represented by the yellow arrow.

Figure \ref{fig:thalamusCircuit}C shows how multiplexer-like relay cells could achieve the orientation transforms shown in figure \ref{fig:thalamusCircuit}A. The colored dots indicate which relay cell inputs would cause the cell to fire under different contextual inputs (yellow) to achieve the three mappings in figure \ref{fig:thalamusCircuit}A. As pictured, the relay cells could perform all possible mappings between input and output. In practice, most of these mappings will never occur; as such, we would not expect to see all relay cells connected to all inputs. In this example, relay cell 04, being at the center of the patch, only needs to relay retinal cell i4. This mixed mapping could explain why some relay cells show near unitary synaptic inputs, while other relay cells receive converging input from several retinal ganglion cells \citep{hammer2015multiple}.

The mapping function could be learned, as opposed to genetically determined. For example, imagine a child looking at an object, and that the child’s head is vertical. This corresponds to the gray bar in A. Next, the child’s head tilts right, corresponding to the green bar. If the active relay cells o2, o4, and o7 stay active after the head rotation, then the relay cells could form synapses with the newly active inputs using Hebbian learning rules. The vestibular system could assist by detecting that the head has rotated, indicating that the active output cells should not change.

\subsubsection{Further Evidence and Predictions for Orientation Transforms in the Thalamus} \label{sec:further_predictions}

As noted, the thalamus could learn to perform the transformations we have proposed. However, as they would take place subcortically, our proposal predicts that they could be applied to any objects that the cortex perceives, even if it has never undergone a specific transformation before - in other words, all of perception would adapt. This is consistent with observations in humans wearing prismatic glasses that invert vision. While wearing these initially causes total disorientation, over the course of around one week, the nervous system adapts enough that subjects can perceive the world relatively unencumbered \citep{Stratton1897, Lillicrap2013}. Learning the necessary transformation in the local structure of the thalamus could explain the rapid adaptation observed across a variety of behavioral and sensory inputs, even in adult humans.

More direct evidence for our proposed role of the thalamus can be found in non-human primate studies. In \citet{Sauvan1999}, rhesus monkeys were mounted in a chair that enabled tilting their heads during single-cell recordings in the primary visual cortex. The authors observed neurons whose response to an oriented edge was invariant to the orientation of the animal. Such neurons were observed as early as V1, suggesting a pre-cortical mechanism enabled the observed response properties.

In \citet{Sauvan1999}, the animal was rotated, but what if the object is rotated? Border ownership cells in V1 have an orientation preference, but they activate uniquely at specific locations on known objects \citep{Zhou2000CodingCortex.}. Our thalamic remapping hypothesis predicts that the response of border ownership cells will be unchanged if a familiar object is rotated. The work of \citet{Franken2021} provides some evidence that such rotation invariance in border ownership cells exists, although demonstrating this prediction conclusively would require evaluating with objects that can be unambiguously rotated, as opposed to the symmetric squares that they used.

\subsection{Hierarchical Connections Model Compositional Objects} \label{sec:composition_objects}
\begin{figure}[!htbp]
    \centering
    \includegraphics[width=\textwidth]{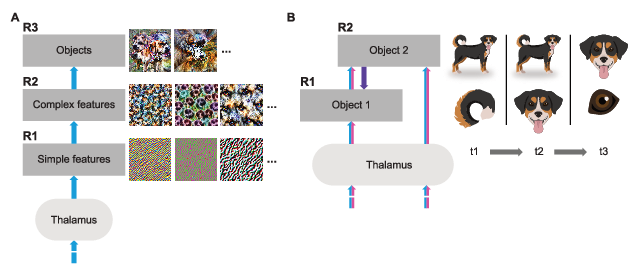}
    \caption{A) The classic view of hierarchical processing in the neocortex. Sensor information enters the neocortex via the thalamus and is then processed in a hierarchy of cortical regions, R1 to R3. Only after several levels of processing is the cortex able to infer complete objects. Neurons at various levels of hierarchy respond preferentially to representations such as edges \citep{Hubel1959ReceptiveCortex, Hubel1962ReceptiveCortex}, contours \citep{Pasupathy1999ResponsesV4, Brincat2004UnderlyingCortex}, and objects \citep{Tanaka1991CodingMonkey, Tanaka1996InferotemporalVision, Quiroga2005InvariantBrain}. Recent forms of this view argue that the visual hierarchy encodes features similar to those found in deep neural networks \citep{Serre2007RobustMechanisms, Yamins2014Performance-optimizedCortex, Schrimpf2018Brain-Score:Brain-Like, Kubilius2019Brain-likeANNs, goh2021multimodal}. Here we show features extracted from a convolutional neural network, adapted from images in \citet{olah2017feature} licensed under CC-BY 4.0. B) Our alternate view of hierarchical processing. Two regions can act in parallel and hierarchically at the same time. All cortical regions receive driving thalamic input (features and movements), allowing them to learn and infer complete objects. The role of the hierarchical connections is to learn compositional structure. That is, the object represented in R1 is a component of the object represented in R2. By attending to different parts of the world, two regions can learn multiple levels of hierarchical composition, for example, a dog has a head (t2), while a head has eyes (t3).}
    \label{fig:Hierarchy old and new}
\end{figure}

Now we turn our attention to the concept of hierarchy, and to the set of hierarchical connections shown in figure \ref{fig:hierarchical_c}. Figure \ref{fig:Hierarchy old and new} is a conceptual illustration of how cortical hierarchy is typically viewed, and our proposal of how it should be viewed.

The conventional view (figure \ref{fig:Hierarchy old and new}A) is that regions learn progressively larger and more complex features as information flows up the hierarchy. The features might be oriented edges \citep{Hubel1959ReceptiveCortex, Hubel1962ReceptiveCortex}, a textured pattern, or an ambiguous contour \citep{Pasupathy1999ResponsesV4, Brincat2004UnderlyingCortex}. Only after several levels of processing is the neocortex able to infer complete objects \citep{Tanaka1991CodingMonkey, Tanaka1996InferotemporalVision, Quiroga2005InvariantBrain}. The feedback connections between regions are often ignored. For example, deep learning neural networks that are trained to recognize images typically use feedforward-only connections during inference \citep{Lecun1998Gradient-basedRecognition, Lecun2015DeepLearning}. The representations in such hierarchical networks are also highly entangled when multiple objects are visible \citep{VonDerMalsburg1999ThePerspective, Locatello2020Object-CentricAttention, zimmermann2023sensitivity}. This contrasts with the object-centric representations that are characteristic of human perception, even for simple perceptual stimuli \citep{spelke1990principles, Treisman1999SolutionsConvergence}. Finally, movement of the sensor during inference is almost always entirely ignored.

The TBT suggests a different interpretation of hierarchy. In figure \ref{fig:Hierarchy old and new} B) we show two regions as in A), but here we emphasize that each region receives driving input from the thalamus, including sensory and movement information. As explained earlier, columns in each region learn structured models of complete objects. The objects modeled and recognized in a cortical column are larger than their input at any point in time. We propose that the role of the hierarchical connections between columns is to learn compositional models, that is, objects that are composed of other objects.

Most of the world is structured this way. For example, a bicycle is composed of a set of other objects, such as wheels, frame, pedals, and seat, arranged relative to each other. Each of these objects, such as a wheel, is itself composed of other objects, such as a tire, rim, valve, and spokes. In another example, words are composed of syllables, which are themselves composed of letters. Finally, the example shown in figure \ref{fig:composition} considers a coffee mug that may have a logo printed on its side. The logo is an object that was previously learned, but in this example, the logo is a component of the coffee mug. Learning compositional objects can occur rapidly, with just a few visual fixations. This tells us that the neocortex does not have to relearn the component objects; the neocortex only has to form links between two existing models. 

Figure \ref{fig:Hierarchy old and new}B shows how two regions might represent a compositional object, such as a dog. R2 represents the parent object, the dog, while R1 represents components of the dog. As the sensor moves, R1 attends to, and recognizes, a series of component objects, such as a head or tail. Each component is associated with a set of locations on the parent object. A moment later, R2 may attend to, and represent, a head, while R1 might represent a component of the head, such as an eye or an ear. Although objects in the world might have many levels of compositional structure, the cortex does not need to represent all these levels at once. As few as two regions can learn multiple levels by successively attending to different subsets of the compositional structure.

According to this proposal, whole objects are represented at all levels of the cortex. Existing studies report the ability to decode complex objects in low-level sensory cortex when measuring from larger neural assemblies \citep{smith2015decoding, Heilbron2020, lanfranchi2025compressed}. While these findings are often interpreted in the context of top-down feedback, we argue that these can be understood in the framework of the TBT, and in particular that low-level cortical columns are themselves capable of encoding entire objects. Supporting evidence for whole objects in primary sensory regions also exists in the form of border ownership cells, neurons whose response properties reflect object-level representations, and yet whose activity emerges in V1 and V2 too quickly to be accounted for by top-down projections alone \citep{Zhou2000CodingCortex., von2023visual}. We predict that by presenting ecologically meaningful objects that are sufficiently small, experimenters would find further evidence of reliably decoding objects in primary sensory cortex.

\subsubsection{Complexities of Compositional Objects} \label{sec:complexities_composite}

\begin{figure}[!htbp]
    \centering
    \includegraphics[width=\textwidth]{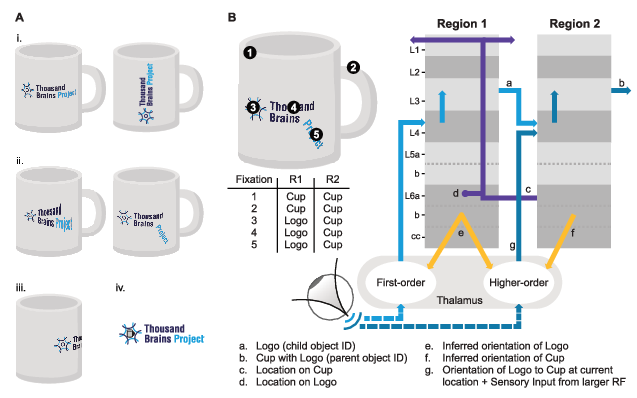}
    \caption{Compositional objects are modeled using hierarchy. A) Learning the relationship of a child object to a parent object is challenging as there are many possible variations. Some of these variations are illustrated here with the placement of a logo on a cup. We are able to quickly learn any of these variations with just a few visual fixations. This suggests there is a common mechanism used by the neocortex to learn almost limitless variations of compositional structure. See text for description. B) Columns in two hierarchically connected regions, R1 and R2, learn compositional objects by attending to a series of physical locations (black numbered circles). Feedforward connections (blue) associate a child object with locations on the parent object. Feedback connections (purple) associate locations on the parent object with locations on the child object. Converging cortico-thalamo-cortical connections (yellow) calculate the orientation of the child object to the parent object. The relationship between child and parent objects is learned on a location-by-location basis, which allows learning unusual arrangements, such as a logo with a bend in it.}
    \label{fig:composition}
\end{figure}

Although we can learn compositional objects quickly and without relearning the component objects, this does not mean that it is simple. Figure \ref{fig:composition}A illustrates some of the complexities involved in learning compositional objects. 

Figure \ref{fig:composition}A i) shows two cups with the same logo. On the first cup, the logo appears horizontal. On the second cup, the logo is vertical (different orientation), and it is slightly larger (different scale). Given examples like this, it is tempting to think that by storing the location, orientation, and scale of a child object relative to the parent object, we can cover all possible arrangements between a child object and a parent object. The two images in figure \ref{fig:composition}A ii) show why that is not true. In one, the logo grows larger toward one end (changing scale), and in the other, the logo bends in the middle (changing orientation and location). In both cases, the logo has been modified from its original form, so it is not possible to simply associate an existing logo model with the cup. Despite these modifications, it is still fast and easy to learn these variations.

Figure \ref{fig:composition}A iii) shows another challenge. The logo was originally learned as a flat two-dimensional object. But when printed on the cup, the logo follows the contours of the three-dimensional cup, first compressing and then disappearing as it wraps around the cup. Surprisingly, we do not perceive the logo as being distorted or truncated. These distortions also do not interfere with our ability to recognize the logo or predict what we will see if the cup is rotated. Figure \ref{fig:composition}A iv) shows a logo with a cup feature, illustrating that in the world, there are no limitations on which objects are parent objects and which objects are child objects. The examples in this figure, while at first confounding, ultimately suggest the mechanism used by the neocortex to represent compositional structure with almost limitless variation.

\subsubsection{Compositional Objects are Learned Location by Location}
The solution to the challenges illustrated in figure \ref{fig:composition}A is that the relationship between child and parent objects is learned on a location-by-location basis. Figure \ref{fig:composition}B shows two columns. One is in region R1, and the other is in a hierarchically higher region R2. The two columns are aligned in sensory space. For example, both columns receive input that is centered on the same point on the retina or the same point on the skin (often at different spatial resolutions). Although the two columns are attending to the same physical location, they are inferring and representing different objects. In this example, R1 represents the logo and R2 represents the cup with the logo.

To learn a compositional object, the columns sense a series of physical locations (black circles labeled 1-5 in figure \ref{fig:composition}). In vision, these would be fixation points between saccades. When the sensor moves over the logo on the cup, each physical location corresponds to a location on the logo in R1 and to the equivalent location on the cup in R2. The feedforward and feedback connections between the two columns allow them to learn the parent/child relationship at each observed location. 

In the feedforward direction, L3 in R1 represents the logo. This becomes the feature sent to L4 in R2. R1 is telling R2 that there is a logo at the current location on the cup. Note that this does not include information about the model of the logo learned in R1, just the object ID. In the feedback direction, L6a in R2 projects back to L6a and then to L1 in R1. The localized synapses in L6a of the R1 column associate the current location on the cup with the coincident location on the logo. R2 is telling R1 that it should be viewing a particular location on the logo. Together, these connections permit the assignment of any set of child objects to any locations on the parent object.

In addition to the localized synapses that form in L6a, the broader projections to L1 enable R2 to bias multiple columns in R1 to perceive the logo. As the sensory inputs to these columns are not guaranteed to be co-aligned, an exact location on the logo cannot be predicted in the neighboring R1 columns. Despite this constraint, providing a broad prediction about the presence of an object has clear benefits for efficient and robust perception.

The asymmetry of the feedforward and feedback connections makes sense. R1 knows it is looking at a logo, but it does not know if that logo is on a cup, a hat, or something else. As such, R1 can't tell R2 what object it should be observing, only that whatever it is looking at should have a logo on it. However, if R2 knows it is looking at the `cup with logo', then it can confidently tell R1 that it should be observing the logo, and that it should expect to be at a particular location on the logo.

What about the example in the figure of a logo with a cup feature \ref{fig:composition}A iv)? Like the head of the dog, which can appear in early or later regions (figure \ref{fig:Hierarchy old and new}B), no object is a priori higher in the hierarchy than another. This is only possible if columns at all levels of the hierarchy are able to learn whole objects, as the TBT proposes.

Hierarchical feedback connections are often described as a prediction, the higher region telling the lower region what feature it should expect. In our proposal, the feedback connections also convey predictions. However, these predictions include the expected object and the expected location on the object.

\subsubsection{Orientation and Scale of Child Objects}
It is not sufficient to just learn which child objects are features at locations on a parent object. A column also needs to learn the orientation and scale of the child objects. Because the orientation and scale of a child object can vary at different locations, as in \ref{fig:composition}A ii, orientation and scale also need to be learned on a location-by-location basis. In addition, the orientation and scale of a child object are relative to the parent. Imagine looking at the cup with the logo, and then rotating the cup in the visual plane. The orientation of the eye to the cup will change and the orientation of the eye to the logo will change, but the relative orientation of the logo to the cup will be constant. It is the relative orientation between child and parent that needs to be learned. Similar logic holds for scale. The size of a parent and child object may change, as when viewing the cup nearby or at a distance, but the relative size of a child object to the parent object will remain constant.

We propose that the calculation of relative orientation between parent and child objects occurs in the thalamus. Recall from figure \ref{fig:cortico_thl_c} that most thalamic nuclei receive converging L6b inputs from hierarchically adjacent cortical columns. In our cup example, the thalamic nucleus that drives R2 receives converging input from L6b in R1 and L6b in R2. These represent the current orientation of the sensor to the logo and the current orientation of the sensor to the cup, respectively. The difference between these two values is the relative orientation of the logo to the cup. It is this value, the relative orientation, that needs to be learned as part of the model of the `cup with logo'.

Earlier, we proposed that the thalamus changes the orientation of features based on the value in L6b. That is true for columns in primary sensory regions. However, for columns that are situated higher in the hierarchy, the thalamus will calculate the relative orientation of the objects in R1 and R2, and use this value to change the orientation of the inputs to R2. In this way, if both the parent and child object rotate together, then the relative orientation between them (and therefore the child-object orientation arriving at the R2 column) will remain fixed. We do not have a detailed proposal at this time for how the thalamus would calculate relative orientation. However, if the feed-forward projection from L6b is able to drive relay cells in higher-order thalamic nuclei, then the same mechanism as described in Section \ref{sec:thalamus_mechanism} could be used to transform the orientation of a child object. There is some evidence that L6b projections to higher-order thalamic nuclei display driving properties \citep{zolnik2026layer}. 

Finally, how does the neocortex calculate and store the relative scale between parent and child objects? Our best guess is that the calculation of relative scale is also performed in the thalamus. A likely mechanism that the thalamus uses to implement scale is one that has been suggested for grid cells, specifically, changing the frequency of one of two oscillators \citep{burgess_2007}, which is sent to the neocortex and changes the rate of path integration. The details of this method are beyond the scope of this paper.

In summary, we propose that hierarchical connections between cortical regions allow columns to learn compositional structure by assigning child objects, represented in a lower region, to parent objects, represented in a higher region. This assignment is done on a location-by-location basis as the sensor moves. The relative orientation and relative scale of the parent and child objects must be calculated, also on a location-by-location basis, which we propose occurs in the thalamus. The complexity of this system is dictated by the variety of compositional models that we can learn.

\section{Discussion}
The primary goal of this paper is to propose how the neocortex models the compositional structure of the world. As part of our proposed solution, we introduced a new interpretation of hierarchical processing in the cortex. We suggest that hierarchically adjacent regions act in parallel as they recognize complete objects, while simultaneously acting hierarchically, learning parent-child relationships between objects. Given this simultaneous, parallel and hierarchical processing, we suggest that the cortex is best described as a heterarchy of regions. We described how just a few cortical regions can learn arbitrarily deep compositional structure with this framework.

The paper also proposes a new theory of the function of thalamic relay cells. As information flows from sensors to cortex and from region to region, there is a need to transform the orientation of features and movements to match the allocentric models learned in cortical columns. We proposed that thalamic relay cells perform this orientation transform. This could explain why the thalamus is intimately connected to all cortical regions, and we proposed a candidate mechanism for how relay cells perform the orientation transform.

Taken together, the theory suggests specific functional roles for the varied long-range cortical-cortical and cortical-thalamic-cortical connections that we described.

\subsection{Significance of the Theory}

Much of our knowledge of the world is compositional. Objects are composed of other objects at varied poses. This is true for language, music, physical objects, and environments. Put simply, the world model learned by the brain is not just a model of things, but is mostly a model of how things are arranged relative to other things in a deep and often complex compositional structure. Computational models of perception often focus on object recognition \citep{DiCarlo2012HowRecognition}, while the powerful compositional aspect of cortical function has received comparatively little attention.

The theory presented in this paper is built upon the sensory-motor premises of the original Thousand Brains Theory. Since most cortical theories do not fully address either sensory-motor processing or compositional models, we believe that the proposals in this paper, along with our previous work, are a significant advance in how we might think about the neocortex, the thalamus, and the mechanisms by which they work.

\subsection{How the Theory Can Be Tested}

There are several novel ideas in this paper supported by empirical data. In developing new theory, we typically begin with a computational requirement, deduce a possible solution, and then search the existing literature for empirical data that could falsify or support the solution. For example, there is a theoretical need to transform the orientation of sensory features from an egocentric reference frame to an object's reference frame to achieve efficient, rotation- and translation-invariant object recognition. We deduced that the thalamus was the most likely place for this transform to occur. However, the solution requires that relay cells form synapses with multiple retinal ganglion cells, which is not what classical relay cell theory suggests. In searching the existing literature, we discovered that researchers had recently found what we predicted \citep{hammer2015multiple}. Another example relates to our proposals on compositional structure. We first assumed that the pose of a child object to a parent object could be specified by three variables: location, orientation, and scale. However, the examples illustrated in figure \ref{fig:composition}B demonstrate that this is insufficient. We then realized that the pose of a child to a parent must be learned at multiple locations, which required location-to-location feedback from L6 in a higher region to L6 in a lower region. These connections should be very specific and localized. Once again, the existing literature contained evidence for such spatially clustered connections \citep{rockland1996collateralized}.

While these experimental findings were unknown to us at the time, theoretical proposals must make novel predictions that are testable. Below, we provide several that we believe have yet to be demonstrated, and which would provide evidence for or against the Thousand Brains Theory 2.0.

\subsubsection{Compositional Structure}

\begin{enumerate}
    \item There is growing evidence that low-level regions, such as V1 and V2, are capable of representing complex objects, not just simple features \citep{lanfranchi2025compressed}. This was predicted by the Thousand Brains Theory. However, our compositional theory predicts that two hierarchically adjacent regions can represent two different objects in a parent-child relationship. The theory predicts this should occur when a subject attends to a child object of a larger parent object.
    \item Learning compositional structure on a location-by-location basis leads to a specific prediction about the feedback connections between two hierarchically connected regions. As shown in figure \ref{fig:hierarchical_c}, these axons start in L6 of the higher region and form narrowly focused synapses in L6 in the lower region, before spreading broadly in L1. The theory predicts that the L6 synapses in the lower region should only exist when the source and destination columns have overlapping receptive fields, such as co-aligned patches of retina.
\end{enumerate}

\subsubsection{Orientation Transforms in the Thalamus}

\begin{enumerate}
    \item There are several ways we might test the hypothesis that relay cells transform orientation. One way is to look at the orientation of the receptive fields of cortical neurons. For example, some cells in V1 respond to edges at preferred orientations. It is assumed that the preferred orientation is relative to the retina. Our theory predicts that in the context of a recognized object, the preferred orientation is relative to the object and not the retina. As noted in Section \ref{sec:further_predictions}, border ownership cells could be used to test this. Border ownership cells have a preferred orientation, but they fire only at certain locations on recognized objects. If a subject is looking at a learned object, and the orientation of the object to the subject changes, we predict that a particular border ownership cell will continue to fire even though the orientation of the edge on the retina has changed. If the same cell is responding without the context of a familiar object, like an oriented grating, then the cell will stop firing as the retina is rotated.
    \item The multiplexer relay cell proposal in figure \ref{fig:thalamusCircuit} requires that most relay cells receive multiple driving inputs. Only one of these inputs should be driving the relay cell at any time. The theory predicts that there will be a synaptic or dendritic mechanism by which some driving inputs are enabled, while most are disabled. This is not how neurons normally behave, but the theory predicts that such a mechanism exists and could be found.
\end{enumerate}

\subsection{Implications for AI}
We believe the theoretical proposals in this work have significant implications for the design of AI systems. Today's leading AI architectures rely on deep learning, a technology that uses multiple levels of artificial neurons and learning via back-propagation of error. This paper describes a different way to learn in a hierarchy. The three key differences are:
\begin{itemize}
    \item Integrating sensor movement with sensed features at each level of the hierarchy to learn structured models of complete objects while interacting with the environment. A deep hierarchy is not required to learn representations that can generalize well.
    \item Using hierarchy to learn how objects in the world are positioned relative to each other, enabling a model of the world that matches the real world's compositional structure.
    \item Learning through local updates within, and between, general processing units (cortical columns) rather than by global calculations of gradients and their backpropagation through an entire network.
\end{itemize}
In the process of developing the TBT 2.0, \citet{thousandbrainsproject2024, Leadholm2025DMC} explored implications of several of its proposals for a learning sensorimotor system. In particular, the equivalent of column-level transformations of incoming and outgoing sensorimotor data was implemented in an architecture known as a Thousand Brains system. This work has demonstrated multiple benefits of these computations, including robustness to novel rotations of objects and enabling principled, model-based movements in the world. The use of reference frames and sensorimotor learning in Thousand Brain systems also confers benefits such as extremely efficient and continual learning, as well as the natural identification of symmetries in the world. These are all major challenges for existing deep learning systems \citep{McCloskey1989CatastrophicProblem, Lake2015Human-levelInduction, ollikka2024comparison, higgins2022symmetry, schneider2024surprising}.

Some of the theoretical ideas in this paper have not yet been tested in computational models, in particular, the implications for learning of compositional objects. We believe these ideas will be critical for future versions of Thousand Brains systems.

\subsection{Conclusion}
In this paper, we have proposed a theory for understanding many of the long-range cortical-cortical and cortical-thalamic connections reported in the literature. Central to understanding these connections is the previous proposal of the Thousand Brains Theory that cortical columns use both sensor and movement information to build structured models. Within this framework, the long-range connections of the neocortex can be understood to fulfill two broad computations: 1) transforming sensor and motor information between egocentric and allocentric reference frames, 2) learning and representing compositional objects. These computations and the connections that support them lead us to propose that the neocortex as a whole is best understood as a heterarchy. While hierarchical connections exist between many regions, the same regions can also operate in parallel via direct input from and output to subcortical structures. As such, the neocortex is not a purely hierarchical system with a complete representation of objects only emerging at the top of the hierarchy. We believe that this framework has significant implications for both understanding the operating principles of the brain and for the design of artificial intelligence.

\section{Funding}

This work was supported by the Gates Foundation (Grant Number: INV-075337). The conclusions and opinions expressed in this work are those of the author(s) alone and shall not be attributed to the Foundation.

This research was financially supported by the Ministry of Trade, Industry and Energy (MOTIE) and Korea Institute for Advancement of Technology (KIAT) through the “International Cooperative R\&D program” (Task No. P0028486).

This work was also supported by the Thousand Brains Project, a non-profit research organization.

This work was also supported by Numenta, a privately held company. Its funding sources are independent investors and venture capitalists.
 
\bibliography{references} 

\section{Appendix}

\subsection{Reference Frames and Grid Cells in Cortical Columns: The Evidence}\label{sec:appendix_grid_cells}

One of the key proposals of the original TBT is that all columns within the neocortex use reference frames to learn structured models of objects. It was further predicted that the reference frames would be mediated by a set of grid-cell-like neurons \citep{Hafting2005MicrostructureCortex} in layer 6 of each column. This set of predictions remains critically important to the TBT 2.0, and it is therefore worth revisiting the origin of these predictions and relevant experimental evidence, including recent findings.

The most fundamental prediction in question is that each column encodes object-centric reference frames that enable path integration. This prediction was derived from the simple observation that the brain makes predictions of sensory input as sensor patches move, observing different parts of learned objects. The existence of reference frames with path integration is a logical prediction of this basic observation. Without such reference frames, it is not clear how columns would be able to predict the sensory inputs they receive following movement.

Why did we propose that cortical columns use grid-cell-like mechanisms for reference frames? Grid cells in the entorhinal cortex exhibit the properties needed by cortical columns. They create reference frames, they perform path integration, and they form new reference frames (re-anchoring) for new environments \citep{Hafting2005MicrostructureCortex, peng2025grid}. It is possible that the neocortex uses a different mechanism for reference frames. The primary reason we propose that the neocortex uses the same, or slightly modified, mechanism as grid cells is an evolutionary one. The grid cell mechanism is complicated. It seems more likely that evolution reused an existing, highly evolved mechanism than discovered a new method to achieve the same effect. If experiments showed that the cortex did not have grid cell-like mechanisms, it would not disprove the major conjectures of the TBT. However, detecting grid cells in the cortex would be strong evidence in support of the TBT.

For a comprehensive discussion of evidence that supported these predictions at the time they were made, and why L6 was proposed as the location of cortical grid cells, interested readers are directed to \citet{Hawkins2017}, \citet{Hawkins2019ANeocortex} and \citet{Lewis2019LocationsCells}. Briefly, experiments conducted in different species and with methodologies ranging from single-cell recordings to fMRI, have identified evidence of grid-cell-like activity in neocortical regions outside of the entorhinal cortex \citep{Doeller2010, jacobs_2013, constantinescu2016organizing, Julian2018HumanGrid, long_2021}. The proposal that layer 6 of neocortex is the likely location for such grid cells is consistent with evidence that it receives motor and movement-related input, such as information about head direction \citep{Velez-Fort_2018}, and projections from motor regions \citep{nelson2013circuit, leinweber2017sensorimotor}. This movement information arrives at L6 even in primary sensory regions such as V1 \citep{leinweber2017sensorimotor}, as predicted by the TBT's proposal that reference frames are utilized throughout the neocortex. Finally, the numerous bidirectional connections between layer 4 and layer 6 within columns \citep{binzegger2004quantitative} match the model first proposed in \citet{Hawkins2017} (see Figure \ref{fig:column}). These findings provide indirect support for the theory, including the specific role proposed for L6.

Since these predictions were made, similar, indirect evidence has accumulated for neocortical grid cells \citep{park2021inferences, liang2024distance, yu2025parietal}. Most notably, \citet{yu2025parietal} recently identified grid-like responses in parietal cortex that encoded the relative positions of visual objects (colored circles) within a larger object. This learned representation of the relative arrangement of features enabled rapid generalization by the human participants on a task. This kind of object-centric reference frame is precisely what the TBT would predict. While promising, it remains an open question what cortical layer is responsible for this grid-like activity, as well as whether the role for grid cells and reference frames is as universal in the neocortex as the TBT claims.

In summary, there remains no conclusive evidence of object-centric reference frames in all cortical columns, or that grid cells in L6 mediate these. However, these continue to be testable predictions, and readers are directed to the discussion in \citet{Lewis2019LocationsCells} for a detailed list of the experimental predictions that remain outstanding. Determining the existence of reference frames in cortical columns faces a number of technical challenges that experiments would need to account for:

\begin{itemize}
    \item Given the sensorimotor nature of the theory, experiments need to be performed while animals are awake and actively interacting with previously learned objects.
    \item While new objects can rapidly be learned in the hippocampal complex, learning happens more slowly in neocortex. This requires designing experiments around ecologically relevant or sufficiently familiar objects.
    \item The TBT predicts that the objects that columns learn will be whole objects of moderate complexity. Objects leveraged in experiments should therefore be sufficiently complex, rather than impoverished stimuli such as sinusoidal gratings.
    \item The TBT predicts that many objects would be encoded in 3D reference frames, yet the response properties of grid cells encoding 3D space are  difficult to interpret, and do not appear as classically "grid-like" \citep{Ginosar2021, Grieves2021}.
    \item Grid cells typically require extended recording periods (on the order of minutes) to identify, which is often incompatible with the natural time scales that an animal interacts with an object \citep{peng2025grid}.
    \item The TBT predicts that the brain leverages distributed neural codes, including for the unique location within a reference frame \citep{Hawkins2016WhyNeocortex, Hawkins2017, Lewis2019LocationsCells}. Identifying such codes requires recording from a large number of cells at a high temporal resolution. The prediction that grid cells are in L6 means such recordings must take place in deep layers, which is particularly challenging for optical techniques \citep{demas2021high}.
\end{itemize}

Fortunately, new techniques are emerging that address many of these challenges, such as the ability to estimate grid cell activity given shorter recording sessions \citep{peng2025grid}, and optical techniques that enable the simultaneous recording of large populations of neurons at greater depths \citep{demas2021high}. With such developments, we expect that more definitive evidence for or against the prediction of cortical reference frames and L6 grid cells will be forthcoming over the coming years. 

\end{document}